\documentclass[preprint,floats,aps,showpacs,epsf]{revtex4}
\usepackage{amssymb}

\usepackage{epsfig}

\newcommand{\be}{\begin{equation}}
\newcommand{\ee}{\end{equation}}
\newcommand{\bea}{\begin{eqnarray}}
\newcommand{\eea}{\end{eqnarray}}

\newcommand{\p}{\partial}
\newcommand{\s}{\sigma}

\newcommand{\la}{\langle}
\newcommand{\ra}{\rangle}
\newcommand{\rd}{\mbox{d}}
\newcommand{\ri}{\mbox{i}}
\newcommand{\re}{\mbox{e}}

\begin{document}
%\draft
\title{Evidence for  the PSL(2$|$2) Wess-Zumino-Novikov-Witten model as a model  for  the plateau transition in Quantum Hall effect: Evaluation of numerical simulations.}

\author{  A.  M. Tsvelik}
\affiliation{
Department of Condensed Matter Physics and Materials Science, Brookhaven National Laboratory, Upton, NY 11973-5000, USA}
\date{\today}

\begin{abstract}
In this paper I revise arguments in favour of  the PSL(2$|$2) Wess-Zumino-Novikov-Witten (WZNW) model as a theory of the plateau transition in Integer Quantum Hall effect. I show that all available numerical data (including the correlation length exponent $\nu$) are consistent with the predictions of such WZNW model with the level $k=8$.   

\end{abstract}

\maketitle
%\narrowtext

%\sloppy
\section{Introduction}

 This paper purports to review the evidence in favour of Wess-Zumino-Novikov-Witten (WZNW) description of the plateau transition in Integer Quantum Hall effect given in \cite{gang}. To avoid the issues related to a possible role of interactions in real Quantum Hall systems I will discuss the plateau transition in a model system, namely in the Chalker-Coddington (CC) network model \cite{CC}. Thus the ``experimental'' data I discuss are numerical data taken from simulations  on the CC networks. 
  
As is well known, all attempts of rigorous derivation of the  critical field theory for the Plateau Transition have remained unsuccessful. The attempts to arrive at such theory by means of educated guess made in  \cite{zirn0} and then in \cite{gang} pointed out to the WZNW model on the PSL(2$|$2) group. However, there have been certain disagreement about precise form of such model which has never been resolved. The papers also did not provide  a value for the correlation length exponent. Below  I argue that the most recent  numerical data are in very good agreement with the predictions of PSL(2$|2)_{k=8}$ WZNW theory. I also identify the relevant operator responsible for the correlation length exponent. 

 At present we have two established models for Integer Quantum Hall effect: the Pruisken-Weidenm\"uller sigma model \cite{pruisken},\cite{W} and the Chalker-Coddington model \cite{CC}. One can arrive to these model descriptions by means of more or less controllable steps from the Schro\"edinger equation description for noninteracting 
 electrons in a disordered potential. However, neither continuous (the sigma model) nor the lattice (the CC model) description  give immediate access to the critical properties. 

 To understand the essence of the problem it is instructive to compare the situation to the quantum critical point in  a half-integer spin Heisenberg antiferromagnet. The CC model can be mapped to  an antiferromagnetic  superspin chain \cite{super}; its  analogue  in this story is  the Heisenberg Hamiltonian:
\be
H = J\sum_n {\bf S}_n{\bf S}_{n +1} \label{Heis}
\ee
Then the analogue of the Pruisken-Weidenm\"uller sigma model would be the O(3) nonlinear sigma model with  the action 
\bea
A = \int \rd\tau\rd x\left\{\frac{1}{2g}(\p_{\mu}{\bf n})^2 + \frac{\ri S}{4}\epsilon_{\mu\nu}\left({\bf n}[\p_{\mu}{\bf n}\times\p_{\nu}{\bf n}]\right)\right\} \label{sigma}
\eea
where ${\bf n}^2 =1$ and $g = 1/2S$. The latter model can be obtained from the former at $S >> 1$ taking the semiclassical limit in the path integral \cite{books}:
\bea
{\bf S}_n = S\left[{\bf m}(x) + (-1)^n{\bf n}(x)\sqrt{1 - {\bf m}^2}\right]
\eea
In that limit ferromagnetic fluctuations described by field ${\bf m}$ are weak and can be  integrated out. In a similar fashion one can derive the Pruisken-Weidenm\"uller sigma model from the CC model \cite{super}. 

Continuing the analogy with the Pruisken-Weidenm\"uller sigma model we can identify $2S$ with the Drude conductivity $\s^{xx}_0$. The second term in (\ref{sigma}) is topological; its contribution to the action is $2\pi \ri S\times$(integer number). The coefficient at the topological term in (\ref{sigma}) should be identified with $\s^{xy}_0$ (strictly speaking we can discuss only the case of $\s^{xy} =1/2$ corresponding to half-integer $S$ or $\s^{xy} =0$ corresponding to integer $S$). 

 Both Pruisken-Weidenm\"uller and O(3) nonlinear sigma models scale towards strong coupling. The topological term gives no contribution to the beta function in {\it any order} in coupling constant $g$ (it gives only nonanalytic contributions). For this reason one would not be able to notice its effect on correlation functions until very large  distances. For the O(3) sigma model the 
corresponding length scale  is 
\be
\xi_{O(3)} \sim a g\exp(\pi/g), ~~ g = 1/2S
\ee
where $a$ is the lattice spacing and for the Pruisken-Weidenm\"uller model it is 
\be
\xi_{Q} \sim \lambda \exp[(4\pi\s_0^{xx})^2]
\ee
where $\lambda$ is the mean free path. One may suspect however, that at large distances  a difference between integer and half-integer spins (or $\s^{xy} =0$ and $\s^{xy} =1/2$) does exist. Indeed, in the first case $\exp[2\pi \ri S] =1$ and there is  no contribution to the partition function, but in the second case the contribution is nontrivial. On that grounds  Haldane made his famous conjecture \cite{haldane} that Heisenberg antiferromagnets with half-integer spins are critical. In the same way following Pruisken \cite{pruisken} we believe that in two dimensions noninteracting disordered electrons with $\s^{xy} =1/2$ are also critical. Of course, in both cases this belief is supported by ample numerical evidence, but for the Heisenberg antiferromagnet extra factors intervene so that for  the Heisenberg magnet we are able to tell the story till the end.  It was conjectured by Polyakov \cite{polyakov} that the critical point, if exists, is  the same as for $S=1/2$ Heisenberg antiferromagnet. This was not that evident because the sigma model derivation is valid only for large $S$. However, later exact solution of  sigma model (\ref{sigma}) was constructed  \cite{fateev} and it was demonstrated that the physical properties thus obtained  indeed interpolate between the O(3) sigma model at small $|x| < \xi$ and $S=1/2$ Heisenberg at large $|x| > \xi$ distances. The critical theory for the $S=1/2$ Heisenberg antiferromagnet is known. The most amazing fact is that this theory is not the O(3) sigma model, but the SU$_1$(2) WZNW model. The order parameter in the critical region is not a three component vector, but a four component matrix $G$. The extra component is the dimerization operator 
\be
d(x) = (-1)^n{\bf S}_n{\bf S}_{n +1}
\ee
which at the critical point has  the same scaling dimension as staggered magnetization ${\bf n}$. So we have
\be
\hat G(x) = \hat I d(x) + \ri{\vec \s}{\bf n}(x)
\ee

 The lessons one takes from the Heisenberg antiferromagnet are  that (i) sigma models with topological terms are not renormalizable and (ii) the resulting critical action may be defined on a manifold greater than the original one.  In that spirit a search for a suitable critical model for the plateau transition was conducted and  the WZNW model on  PSL$(2|2)$ group was suggested as a candidate\cite{zirn0},\cite{gang}:
\bea
A = \frac{1}{2g}\int \rd^2x \mbox{Str}(\p_{\mu}G^{-1}\p_{\mu}G) + \ri k\Gamma[G]\label{wznw}
\eea
Here $G$ is a matrix from PSL(2$|$2) group, $\Gamma[G]$ is the Wess-Zumino term (not to be confused with the topological one present in the original sigma model), $k$ is an integer. In \cite{zirn0} Zirnbauer argued for $k=1$, in \cite{gang} it was suggested that  $g = 1/k$; in the latter  case the model  has an extended symmetry ${ PSL(2|2)}_R\times{ PSL(2|2)}_L$. 

 The standard situation with  WZNW models is that it is critical only at $g=1/k$ and a deviation of $g$ from $1/k$ generates an irrelevant perturbation. The corresponding  RG equation  is
\be
\frac{\rd g}{\rd\ln(\Lambda/k)} = -a(g - 1/k)
\ee
with $a \sim C_{adj} > 0$, where $C_{adj}$ is the quadratic Casimir in the adjoint representation of the group. Since for PSL(n$|$n) group this Casimir is zero, it was argued that (\ref{wznw}) was critical for any value of $g$ \cite{zirn0} though  the extended (Kac-Moody) symmetry exists only at $g = 1/k$. 
It was also suggested  in \cite{zirn0} that there was no Kac-Moody  symmetry at the plateau transition. One reason for that assumption was that the critical coupling $g$ was identified with the critical value of the conductivity:
\be
g = 1/\pi\sigma^{xx} = 2/\pi \label{critg}
\ee
 which is not integer. Since the Kac-Moody symmetry requires $1/g = k$ to be integer, one is forced to conclude that it is not there. The argument however hangs on the identication (\ref{critg}). However, as was pointed out in \cite{gang}, since the critical action is not the original sigma model, such assumption is really difficult to justify. To this it should be added that since  at the critical point the conductivity has has O(1) mesoscopic fluctuations, it is not quite clear what to mean by $\s_{xx}$. To talk about its  value one must specify how  one averages over the conductance distribution.

 In any case, in the absence of Kac-Moody symmetry we know nothing about critical properties of model (\ref{wznw}). Therefore we cannot compare the available numerical evidence with theoretical predictions which do not exist. Meanwhile for the Kac-Moody WZNW model we know quite a lot and  can compare. Below I will argue that  such  comparison leads to quite a satisfactory  agreement for one particular value $k =8$. At present I have no clue why such value of $k$ may be  chosen by the scaling. In my defence I can say that properties of WZNW models on supermanifolds are not well studied and future research in this direction is required to establish the soundness of my conjecture.

\section{Operators, scaling dimensions and the numerical data}

 The available numerical evidence  essentially consists of four sets of data. Each set contains information related to certain universal properties of the model. 
\begin{itemize}
\item
For eigenstates away from the critical energy $E_c$ we know the localization (correlation) length exponent
\be
\xi(E) \sim (E - E_c)^{-\nu}, ~~ \nu \approx 2.3-2.35
\ee
\item
From \cite{zirn1},\cite{zirn2} we know the statistical distribution of the two-point conductances $P[T(x,y)]$.
\item
From \cite{mirlin} we know the statistical distribution of the wave functions (participation ratios). We also have an estimate of the scaling dimension of the first irrelevant operator $d = 2.38 \pm 0.04$ \cite{mirlin},\cite{bodo}.  
\end{itemize}

 Few words about PSL(2$|$2) group are in order.  This group consists of 4$\times$4 supermatices with unit superdeterminant. Matrix $G$ from this  group can be parameterized as follows:
\bea
G = \re^{\varphi}\left(
\begin{array}{cc} 
I & 0\\
\theta & I
\end{array}
\right)\left(
\begin{array}{cc} 
A & 0\\
0 & B
\end{array}
\right)\left(
\begin{array}{cc} 
I & \bar\theta\\
0 & I
\end{array}
\right)
\eea
where $\theta, \bar\theta$ are fermionic $4\times 4$ matrices, $A$ is a matrix from $H^3_+ = SL(2,C)/SU(2)$ and $B$ is SU(2) matrix. 

It turns out that  field $\varphi$ does not enter into action (\ref{wznw}). Therefore  the PSL(2$|$2) model can be understood as SL(2$|$2), but with a certain restriction on the Hilbert space. Namely, one has to consider only operators which are invariant under {\it local} multiplication of field $G$ by a number, that is fields of the general form 
\bea
Q_X = G X G^{-1}, ~~ \mbox{Sdet}X =1 \label{Q}
\eea
with $X$ being constant matrices.

 The axioms of the conformal field theory state that one can introduce a basis in the space of fields so that every field $\Phi(x,y)$ in the theory can be decomposed into a sum of fields (operators) from this basis. The operators which compose the basis can be grouped into ``conformal towers''. Each tower consists of a multiplet of so-called primary fields and their descendants. The latter ones are fields generated from the primaries by generators of conformal transformations. Conformal dimensions of descendants differ from dimensions of their primaries by positive integer numbers. For this reason  descendants  are less relevant.

 In WZNW models primary fields compose multiplets such that fields within each multiplet transform into each other under  action of the group. The generators of group transformations are currents composing a Kac-Moody algebra. 
The scaling dimensions of the primary fields in the PSL(2$|$2) WZNW model are \cite{schomerus}
\bea
h^{(j_F,j_B)} = \bar h^{(j_F,j_B)} = \frac{1}{k}[j_F(j_F + 1) - j_B(j_B +1)] \label{dims}
\eea
Here $j_F$ characterizing  the compact sector run through the discrete series $j_F = 0, 1, ... k/2 -1$ (Eq.(\ref{Q}) excludes half-integer $j_F$). The nagular moment eigenvalue $j_B$ may be either discrete $j_2 = -q$ ($q$ is a positive real number $q \leq (k+1)/2$) or continuous $j_B = -1/2 + \ri p/2$. In the first case 
the corresponding representations of su(1,1) are $(+,j_B)$ and $(-,j_B)$ which denote representations with lowest/highest weight with eigenvalues of the Cartan operator $K_1^0$ being $q + n$ ($n=0,1,2,...$) and $-q - n$ respectively. In the latter case the representations have neither highest nor lowest weight states. Eigenvalues of $K_1^0$ take values on $\alpha +n$ where $0 < \alpha < 1$, but the Casimir is independent on $\alpha$. 

 If PSL(2$|$2) WZNW indeed describes the critical point of CC model there must exist {\it integer} $k$ such that dimensions (\ref{dims}) describe power asymptotics  of various correlation functions. This requirement stands  for any WZNW model defined on a group containing compact submanifolds (such as SU(2) subgroup for the PSL(2$|$2)).  

 Let us now return to the  numerical evidence and see whether it supports (\ref{dims}) with integer $k$. It was suggested in \cite{gang} that the operators corresponding to the $q$-th power of local densities of states (DOS) have $j_F=0, j_B = -q$ such that  
\be
h_q = \bar h_q = \frac{q(1-q)}{k} \label{hdos}
\ee
This hypothesis was checked against numerical calculations \cite{mirlin}. This work also reported that numerical simulations on CC networks are plagued by strong size effects and to get valuable estimates one should  consider really large samples. Simulations conducted on large samples confirmed the parabolic spectrum (\ref{hdos}) with  $2/k = 0.26 \pm 0.003$. The parabolicity of the spectrum is all important for WZNW model interpretation. As far as the numerical  value of $k$ is concerned, it is rather close  to integer value $k =8$.

 Now let us discuss the localization length exponent $\nu$. In the sigma model the corresponding perturbation is generated by deviation of $\s_{xy}$ from its critical value 1/2. Since $\s_{xy}$ stands at the topological term and the latter  term gives nontrivial contributions only when  the compact sector of the theory is involved, the perturbation must have nonzero $j_F$. As I mentioned above, $j_F$ must be integer. The operator with  $j_F =2, j_B = -1/2 + \ri p/2$ (thus  it is not a single operator, but rather a continuum of fields) does the job. The corresponding scaling dimensions are 
\be
d(p) = \frac{25 + p^2}{16}
\ee
As far as physical quantities are concerned, their behavior is dominated by $d(p0$ at $p << 1$ (see the discussion in Section III). The minimal scaling dimension of the correlation function $\nu = 1/[2 - d(0)] = 16/7 \approx 2.29$ differs by 2 percent from the accepted value 2.35. The fact that the perturbation consists of the entire continuum of operators must generate strong size effects. Below I will return to this matter and discuss it in more detail.  

 It well may be that the first decendant of the operator with $j_F =1,j_B = -1/2 + \ri p/2$, with scaling dimension 
\bea
d_{irr}(p) = 2 + \frac{9 + p^2}{16}   \label{irr} 
\eea
also fits in the picture. If we believe that   the finite size corrections to the wave functions in \cite{mirlin},\cite{bodo} are dominated by the region of small $p$ than it gives for the observed scaling dimension the value $d_{irr} -2 \approx 0.45$ which is not that different from the value $0.38 \pm 0.04$ given in \cite{mirlin},\cite{bodo}. 

 As we see, $k=8$ works rather well on two pieces of numerics itemized above. In that light I suggest to look again at  earlier numerical experiments of \cite{zirn1},\cite{zirn2}. In \cite{zirn1} it was suggested that an average of $q$-th power of two-point conductance measured between points $(x,y)$ and $(0,0)$ on a cylinder is given by 
\bea
\la T^q(x,y)\ra = \int_0^{\infty} \rd \mu(p) |C(q,p)|^2|\pi/W\sinh[\pi(x + \ri y)/W]|^{-4h(p)}\label{T}
\eea
where $W$ is the circumference of the cylinder, $h(p)$ corresponds to $j_F=0, 
j_B = -1/2 +\ri p/2$, 
\[
\rd \mu(p) = \rd p \frac{p }{2}\tanh(\pi p/2)
\]
is the Plancherel measure for the continuous series and $C(q,p)$ is the Clebsh-Gordan coefficient. Comparing numerics with (\ref{T}) the  authors of \cite{zirn1}  found that  
\begin{itemize}
\item
The conductance does scale with 
\[
\zeta = |\pi/W\sinh[\pi(x + \ri y)/W]| 
\]
which strongly supports the conformal invariance,
\item
The scaling dimension is parabolic
\[
2h(p) = (1 + p^2)X_t/8
\]
\item
It was found that $X_t = 0.63 \pm 0.03$.
\end{itemize}
 The numerical value of $X_t$ was judged to be close to $2/\pi$ which excluded integer $k$ and worked as an argument against the WZNW model as a possible candidate for the plateau transition. However, the later improved numerical calculations reported in \cite{zirn2} found $X_t =0.54$ and $X_t = 0.57$ which is much closer to the desired value 0.5. It is also likely  that a limited accuracy of the numerical calculations is not the only source of deviations and there are systematic errors. The point is that the formulae for the powers of conductance suggested of \cite{zirn1},\cite{zirn2}  interpret  the fusion factor $C(q,p)$  as 
the Clebsch-Gordan coefficient of the GL(2$|$2) group. Such choice  is an approximation which can be justified only in the limit $k \rightarrow \infty$. For quantum field theory structure constants characterizing a fusion of two operators do not coincide with Clebsch-Gordan coefficients of  the group theory. 

 Let me  illustrate this argument by an example. Consider some well familiar group, say,  the SU(2) one. The representations of this group are labeled by spin $j$. A tensor product of two operators (matrices) belonging to representations $j_1,j_2$ can be decomposed into the sum
\bea
\Phi^{(j_1)}\Phi^{(j_2)} = \sum_{j_3 = |j_1-j_2|}^{j_1 + j_2}C^{(j_1,j_2)}_{j_3}\Phi^{(j_3)} \label{al}
 \eea
Now let us consider WZNW model on the same group. The operators carry the group indices and transform according to representations of the group:
\[
\Phi^{j}_{m,m'}(z,\bar z)
\]
The crucial difference is that now they depend on a space point, so they are not simply matrices from SU(2) group, but fields. Studying correlation functions of the corresponding WZNW model 
instead of (\ref{al}) one arrives to more complicated fusion rules:
\bea
&& \Phi^{(j_1)}(z_1,\bar z_1)\Phi^{(j_2)}(z_2,\bar z_2) = \\
&&\sum_{j_3 = |j_1-j_2|}^{j_1 + j_2}|z_{12}|^{(d_3 - d_1 -d_2)}{\cal C}^{(j_1,j_2)}_{j_3}[\Phi^{(j_3)}(z_2,\bar z_2) + z_{12}^n{\bar  z_{12}}^mA_{n,m}\Phi^{(j_3;n,m)}(z_2,\bar z_2)]\nonumber
\eea
where $\Phi^{(j_3,n)}$ are operators with conformal dimensions $h(j_3) +n,h(j_3) + m$ (decendants) and $A_{nm}$ are numerical coefficients. The quantum Clebsch-Gordan coefficients ${\cal C}$ do not coincide with the group coefficients  except for $k \rightarrow \infty$\cite{difranc} .

\section{Predictions for the off-critical scaling}

I suggest that at $E \neq E_c$ the effective action has the form
\bea
S = S^* + a(E - E_c)\int \rd z\rd\bar z \int \rd \mu(p) \mbox{Str}\left[\Phi^{(j_F=2,j_B = -1/2 + \ri p/2)}(z,\bar z)\right]
\eea
where $S^*$ is the critical action. 
As a consequence the corrections to physical quantities in a system of size $L$ will include integrals as
\bea
\tau \int \rd \mu(p) L^{(2 - d(p))} =  \tau L^{2 - d(0)}\int \rd p p \tanh(\pi p/2)\exp\left[- \frac{\ln L}{16}p^2\right] \label{int}
\eea
From this expression one can conclude that the dominant contribution to scaling will come from $d(0)$, but there will be significant logarithmic corrections. These corrections will be a source of errors in one-parameter scaling fits of numerical data obtained for finite samples. Since logarithm is a slow function, fits made in the limited range of sample sizes will produce a perfect illusion of a single power law for a given physical quantity.  The corrections to power laws will survive even for very large samples when  the integral (\ref{int}) is dominated by $p << 1$ (this corresponds to  $L > \re^{16} \sim 10^4$). Indeed,  the estimate of (\ref{int}) shows  us that at large $L$ the parameter $\tau = a(E-E_c)$ enters in combination
\bea
\tau L^{7/16}[\ln L]^{-3/2}
\eea
so that the correlation length behaves as 
\bea
\xi \sim \tau^{16/7}[\ln \tau]^{24/7}
\eea
 The described effects  set a very tough standard for the size of the critical region and may explain large magnitude of the size effects observed in \cite{mirlin}.

\section{Conclusions and Acknowledgements}

 The review of numerical results on Plateau Transition in the CC model demonstrates that PSL$(2|2)$ WZNW model with k=8 stands as a very good candidate to model this transition. It goes without saying that this identification leaves many questions unanswered. One may wonder, for instance, what makes k=8 so special or why the correlation length exponent is controlled by the operator with $j_F =2$ and not with $j_F =1$. All these questions can be answered only when detailed information about operator expansion in PSL WZNW models will be available.

 As a way forward I can suggest two things. First, one should study the operator algebra for the  PSL$(2|2)$ WZNW models and see whether it possesses some peculiar properties which allow selection of certain values of $k$. From this algebra one will also be able to extract quantum Clebsch-Gordan coefficients to use in (\ref{T}). In this way we will obtain more accurate theoretical prediction for the distribution of conductances. Second, one may undertake to study numerically deviations from a simple scaling predicted in the last Section. In this context it should be emphasised that continuous spectra of scaling dimensions, being a generic feature of critical sigma models on noncompact manifolds, is not a unique feature of the PSL(2$|$2) WZNW theory. 

 This work was supported by US DOE under contract number DE-AC02 -98 CH 10886. I am grateful for hospitality to Abdus Salam ICTP where this work was performed.  I also gratefully acknowledge valuable conversations with V. Kravtsov, M. Skvortsov, A. Mirlin  and C. Mudry.


\begin{thebibliography}{99}


\bibitem{gang} M. J. Bhaseen, I. I. Kogan, O. A. Solovyev, N. Taniguchi and A. M. Tsvelik, Nucl. Phys. B{\bf 580}, 688 (2000).
\bibitem{CC} J. T. Chalker and P. D. Coddington, J. Phys. C{\bf 21}, 2665 (1988)\bibitem{zirn0} M. R. Zirnbauer, cond-mat/09905054.
\bibitem{pruisken} A. M. M.  Pruisken, Nucl. Phys. B{\bf 235}, 277 (1984).
\bibitem{W} H. A. Weidenm\"uller, Nucl. Phys. B{\bf 290}, 87 (1987).
\bibitem{super} M. R. Zirnbauer, Ann. Physik {\bf 3}, 513 (1994). 
\bibitem{books} E. Fradkin, {\it Field Theories in Condensed Matter Systems}, Addison-Wesley, New York, 1991.; A. M. Tsvelik {\it Quantum Field Theory in Condensed Matter Physics}, Cambridge University Press, 2003. 
\bibitem{haldane} F. D. M. Haldane, Phys. Lett A{\bf 93}, 464 (1983); Phys. Rev. Lett. {\bf 50}, 1153 (1983); J. Appl. Phys. {\bf 57}, 3359 (1985). 
\bibitem{polyakov} A. M. Polyakov, unpublished. 
\bibitem{fateev} V. A. Fateev and A. B. Zamolodchikov, Phys. Lett B{\bf 271}, 91 (1991).
 \bibitem{zirn1} M. Janssen, M. Metzler and  M. R. Zirnbauer, Phys. Rev. B{\bf 59}, 15836 (1999).
\bibitem{zirn2} R. Klesse and M. R. Zirnbauer, Phys. Rev. Lett. {\bf 86}, 2094 (2001).
\bibitem{mirlin} F. Evers, A. Mildenberger and A. D. Mirlin, Phys. Rev. B{\bf 64}, 241303(R) (2001).
\bibitem{bodo} B. Huckestein, Phys. Rev. Lett. {\bf 72}, 1080 (1994).  
\bibitem{schomerus} G. G\"otz, T. Quella and V. Schomerus, hep-th/0610070.
\bibitem{difranc} One can consult the  book {\it Conformal Field Theory} by 
Ph. Di Franchesco, P. Mathieu and D. Senechal, Springer, Berlin, 1999. 
%\bibitem{leclair} B. Gerganov, A. LeClair and M. Moriconi, Phys. Rev. Lett. {\b%f 86}, 4753 (2001).
%\bibitem{mudry} S. Ryu, C. Mudry, A. Furusaki and A. W. W. Ludwig, cond-mat/061%0598.  
\end{thebibliography}
\end{document}